\documentclass{jetpl}
\twocolumn
\title{Particle decay in de Sitter spacetime via quantum tunneling}

\rtitle{Particle decay in de Sitter spacetime via quantum tunneling}

\sodtitle{Particle decay in de Sitter spacetime via quantum tunneling}

\author{G.E. Volovik
 $^{\#}$\/\thanks{
volovik@boojum.hut.fi} 
}

\rauthor{G.E.Volovik}

\sodauthor{Volovik}

\address{Low Temperature Laboratory, Helsinki University of
Technology, P.O.Box 5100, FIN-02015, HUT, Finland
\\
 Landau Institute for Theoretical Physics RAS, Kosygina 2,
119334 Moscow, Russia}

\dates{June 2, 2009}{*}

\abstract{
The quantum tunneling process of decay of the composite particle in the de Sitter vacuum  looks as thermal radiation with the effective temperature twice larger than the  Hawking temperature associated with the cosmological horizon. 
 }

\begin{document}

\maketitle


\section{Introduction}

Both  the  black hole horizon and the cosmological horizon are described by 
the so-called ``fluid'' metric which is characterized by the ``velocity'' field ${\bf v}$ \cite{Visser2005}:
\begin{equation}
ds^2=g_{\mu\nu}x^\mu x_\nu= -dt^2 + \left(d {\bf r}- {\bf v}dt\right)^2~,
\label{FluidMetric1}
\end{equation}
where we used the units with $c=1$.
The de Sitter spacetime is characterized by the radial velocity field
\begin{equation}
{\bf v}({\bf r})=v(r)\hat{\bf r} ~~,~~v(r) =Hr=  \frac{r}{r_H} ~,
\label{FRWfluid}
\end{equation}
where $r_H=1/H$ is the radius of cosmological horizon.
The  ``fluid'' metric for black hole  at the end of the gravitational collapse is the Painleve-Gullstrand metric
\cite{Painleve}, which corresponds to the radial flow field in the form:
   \begin{equation}
 {\bf v}({\bf r})=v(r)\hat{\bf r} ~~,~~v(r)= -\sqrt{\frac{r_H}{r}}~~,
\label{Schwarzschild}
\end{equation}
where $r_H$ is the radius of the black hole horizon.
The  ``fluid'' metric is best suited for the derivation of the Hawking radiation using the semiclassical tunneling picture \cite{Volovik1999a,ParikhWilczek}, because  this metric is stationary and thus the energy is well defined. The classical energy  spectrum of a particle with mass $m$ in the ``fluid''  spacetime is given by
\begin{equation}
E({\bf p},{\bf r})   = \sqrt{m^2 + p^2}+  {\bf p}\cdot {\bf v}({\bf r}) ~,
\label{Spectrum}
\end{equation}
where the first term is the spectrum in the frame  ``comoving with the vacuum'', while the last term plays the role of the Doppler frequency shift. 
The tunneling probability is obtained from the imaginary part of the  action along the semiclassical trajectory
\begin{equation}
  w=w_0 \exp(-2{\bf Im}~S) 
~.
\label{TunnelingProbability}
\end{equation}
The  radial trajectory $p_r(r)$ is obtained from the energy conservation along the trajectory:
\begin{equation}
E(p_r,r)   = \sqrt{m^2 + p_r^2}+ p_r v(r)= E ~,
\label{SpectrumTraj}
\end{equation}
which gives the tunneling exponent
\begin{equation}
2{\bf Im}~S=2{\bf Im}\int dr ~p_r(r) = \frac{2\pi E} {|dv/dr|_{r=r_H}}~.
\label{Action}
\end{equation}
The quantum tunneling thus simulates the thermal  radiation from the horizon with Hawking temperature
\begin{equation}
T_{\rm H}= \frac{\hbar}{2\pi } \Big |\frac{dv} {dr}\Big |_{r=r_H}~.
\label{HawkingT}
\end{equation}
For the de Sitter Universe with its $v(r)=Hr$, the corresponding temperature
would be
\begin{equation}
T_{\rm H}^{\rm dS}= \frac{\hbar H}{2\pi }~.
\label{HawkingTDS}
\end{equation}
However, in this semiclassical description the prefactor $w_0$ in \eqref{TunnelingProbability} remains unknown, and there are arguments that the symmetry of  the de Sitter background nullifies the prefactor
\cite{Volovik2009,volovik1}. For a discussion of the controversies
concerning the stability of de-Sitter vacuum towards the Hawking radiation, see e.g.
Refs.~\cite{Starobinskii1979,StarobinskyYokoyama1994,GarrigaTanaka2007,
TsamisWoodard2007,Polyakov2008,Busch2008}.  
  
While the de Sitter vacuum may be stable, the particles living in the de Sitter environment are certainly not
 \cite{Nachtmann1967}. This is because the mass of particle is not well defined in the de Sitter background.  Calculations of the decay rate of the composite particles have been done in 
 Refs. \cite{Bros2009,Bros2008,Bros2006}. Let us stress, that contrary to conclusion made in Ref. \cite{AkhmedovBuividovichSingleton2009}, we argue that the possibility of massive free falling particles to radiate other massive particles does not mean that the de Sitter space cannot exist eternally. 
 In the presence of  an external body (detector or composite particle), the radiation occurs which takes the energy from the body. But the pure de Sitter vacuum (i.e. without any impurity) may be stable.
 
 Here we use the semiclassical tunneling picture for the calculation of the decay rate and make comparison with the Hawking radiation.
  
  \section{Ionization rate and Hawking temperature}
\label{DdSb}

We consider two examples of the radiation caused by the presence of external object in the de Sitter vacuum: ionization of an atom caused by the de Sitter expansion discussed in \cite{Volovik2009}, and the decay of the composite particle into two particles  in the de Sitter background discussed in 
Refs. \cite{Bros2009,Bros2008,Bros2006}.
The atom (or any other composite or massive particle) plays two roles: it serves as the detector of radiation; and it  violates the de Sitter symmetry and provides the nonzero matrix element for the radiation, since as we argue the pure de Sitter vacuum is not radiating due its symmetry. 

Let us start with an atom \cite{Volovik2009}, which is at rest in the comoving reference frame. In the reference frame of the atom its position is at the origin, $r=0$. The electron  bounded to an atom   absorbs the energy from the gravitational field of the de Sitter background, which is sufficient to escape from the electric potential barrier that originally confined it.
If the electron is originally 
sitting at the energy level $E_n$, then the  ionization potential $\epsilon_0=-E_n$. If the  ionization potential is much smaller than the electron mass,  $\epsilon_0\ll m$ , one can use the non-relativistic quantum mechanics to estimate the tunneling rate through the barrier. 
The corresponding radial trajectory $p_r(r)$ is obtained from the classical equation   $E(p_r,r)= -\epsilon_0$,
which in the non-relativistic approximation reads 
 \begin{equation}
 -\epsilon_0=\frac{p_r^2(r)}{2m} +  p_r(r)Hr~.
\label{RadialTrajDS}
\end{equation}
Here $p_r$ is the radial momentum of electron, and the last term 
is the  Doppler shift $p_rv(r)$ in Eq.  \eqref{Spectrum} provided by the  de Sitter expansion   \eqref{FRWfluid}.  This gives the following radial trajectory of electron:
\begin{equation}
p_r(r)=-mHr + \sqrt{m^2H^2r^2 -2m\epsilon_0}~.
\label{RadialTrajDS2}
\end{equation}
The sign in front of the square root is chosen such that it corresponds to the flux from the center, i.e. the radial velocity of the particle $u_r=dE/dp_r= p_r/m +Hr$ is positive in the classically allowed region  $r>r_0$, where 
\begin{equation}
r_0^2=\frac{2\epsilon_0}{mH^2}~.
\label{RegionBarrier}
\end{equation}
The momentum $p_r$ is imaginary in the classically forbidden region $0<r<r_0$, 
which demonstrates that there is an energy barrier between the position of the electron in the atom, i.e. at $r=0$, and the position of the free electron with the same energy at $r=r_0$.
Since we assume that $\epsilon_0\ll m$, one has $r_0\ll r_H=1/H$, which means that tunneling occurs well within the horizon. We also assume that $H\ll \epsilon_0 (\epsilon_0/m)^{1/2} \alpha^{-1}$, which
allows us to neglect the region close to the origin where the contribution of the Coulomb potential $-\alpha/r$ to Eq.(\ref{RadialTrajDS}) is important.

The imaginary part of the action
\begin{equation}
{\bf Im}\int dr ~p_r(r)=mH\int_0^{r_0}dr \sqrt{r_0^2-r^2}=
\frac{\pi\epsilon_0}{2H}~,
\label{IonizationExponent}
\end{equation}
gives the probability of ionization 
\begin{equation}
w\propto \exp(-2{\bf Im}~S) 
=\exp\left(-\frac{\pi\epsilon_0}{H}\right)~.
\label{IonizationProbability}
\end{equation}
The quantum tunneling of the electron in the gravitational field of the de Sitter spacetime thus simulates the thermal activation of an atom by a heat bath with effective temperature $T$, which is twice larger than the corresponding Hawking temperature  in \eqref{HawkingTDS}:
\begin{equation}
w\propto  \exp\left(-\frac{\epsilon_0}{T}\right)~~,~~T=\frac{\hbar H}{\pi}=2T_{\rm H}^{\rm dS}~,
\label{ActivationT}
\end{equation}

 \section{Decay rate of composite particle and Hawking temperature}
 
The same result is obtained in Refs. 
\cite{Bros2009,Bros2008,Bros2006}, whose authors considered the decay of a composite particle with mass $m_0$ into two particles, each  with mass  $m_1 > m_0/2$. Such decay is energetically forbidden in the Minkowski spacetime, but is allowed in the de Sitter background.
It is instructive to derive the results of Refs. \cite{Bros2009,Bros2008,Bros2006} using also the semiclassical tunneling picture. The trajectory of each of the two particles with mass $m_1$ moving in the radial direction from the origin at $r=0$ is obtained from equation
\begin{equation}
E(p_r,r) = \sqrt{p_r^2 + m_1^2} + p_rHr= \frac{m_0}{2}~.
\label{ParticleTraj1}
\end{equation}
We took into account that each of the two particles carries the one half of the energy of the original particle, i.e. $E=m_0/2$.  The momentum along the trajectory is
\begin{equation}
p_r(r) = \frac{1}{1-H^2r^2} \left[- \frac{m_0}{2}Hr +\sqrt{ \frac{m_0^2}{4}-m_1^2+ m_1^2H^2r^2}\right].
\label{ParticleTraj2}
\end{equation}
Here again we choose the sign in front of the square root, which in the classically allowed region at $r>r_0$, where $r_0=\sqrt{1-m_0^2/4m_1^2}/H$, corresponds to the classical motion from the center. The momentum is imaginary in the classically forbidden region $r<r_0$, which gives the imaginary contribution to the action:
\begin{equation}
{\bf Im}\int dr ~p_r(r)=\frac{m_1}{H}\int_0^{r_0}dr \frac{\sqrt{r_0^2-r^2}}{r_H^2-r^2} = \frac{\pi}{4}(2m_1-m_0).
\label{DecayExponent}
\end{equation}
where as before $r_H=1/H$ is the position of the de Sitter horizon. We must take into account that due to momentum conservation, the two particles tunnel simultaneously in opposite direction (which is called co-tunneling). This adds extra factor two in the exponent. As a result one obtains the decay rate:
\begin{equation}
w\propto\exp(-4{\bf Im}~S) =\exp\left(-\frac{\pi(2m_1 -m_0)}{H}\right)~.
\label{DecayRate2}
\end{equation}
This looks as thermal activation by a heat bath with the temperature  which is again twice larger than
the corresponding Hawking temperature in \eqref{HawkingTDS}:  
\begin{equation}
w\propto \exp\left(-\frac{\pi\Delta m}{H}\right)= \exp\left(-\frac{\Delta m}{T}\right)~~,~~T=\frac{\hbar H}{\pi}=2T_{\rm H}^{\rm dS}~.
\label{DecayRate}
\end{equation}
Here $\Delta m$ is the mass deficit. It is the analog of  the  ionization potential  $\epsilon_0$ in \eqref{IonizationProbability}. In the case of the decay of the particle with mass $m_0$ into two  particles with masses $m_1$, the mass deficit is
$\Delta m=2m_1-m_0>0$.

\section{Discussion}

The decay rate calculated using the semiclassical method reproduces exact result
obtained in Refs.~\cite{Bros2009,Bros2008,Bros2006} (note that there is misprint in Eq.(16) of Ref.~\cite{Bros2006}:  the factor $\pi$ has  been omitted in the  exponent). Both approaches demonstrate that the effective temperature which characterizes the decay rate of composite particles in de Sitter space is twice larger than the Hawking temperature of the de Sitter horizon. The controversies concerning the factor of 2 for the Hawking and Unruh temperatures can be found in Refs. \cite{AkhmedovaPillingGillSingleton2008,Pilling2008,Akhmedova2-2008,Akhmedov2008} and references therein. However, the same semiclassical method applied to the black hole radiation  
\cite{Volovik1999a} gives rise to the correct factor  in the Hawking temperature. In the case of the Unruh effect \cite{Unruh1976}, the tunneling approach is different because of the time dependent potential   \cite{Volovik1992}, but it also gives the correct factor for the Unruh temperature.

It is important that the effective temperature (\ref{ActivationT}) has nothing to do with 
the existence of the cosmological horizon, since both for the atom and for the decaying particle the energy barrier 
is situated  within the horizon: $r_0<r_H=1/H$. Moreover, the residue of the pole at $r=r_H$ in Eq. \eqref{ParticleTraj2} vanishes.
That is why the possible subtleties, which may influence the semiclassical 
tunneling approach in the presence of horizon 
and restore the `correct' factor \cite{AkhmedovaPillingGillSingleton2008,Pilling2008,Akhmedova2-2008,Akhmedov2008},
are irrelevant here. The extra factor of 2 appears in some calculations of the Hawking temperature, when people use the Schwarzschild static coordinates in the tunneling method (see also discussion in Ref.  \cite{Ya-PengHu2009}). The Schwarzschild coordinates are not well suited for calculations, since they have coordinate singularity at the horizon and do not describe the interior of the black hole. 
The Painlev\'e-Gullstrand  ``fluid'' metric  used in Refs. \cite{Volovik1999a,ParikhWilczek} and in the present paper does not suffer from such drawbacks. That is why the extra factor 2 which appears in the decay of composite particle is not the artefact of the wrong coordinates.

It is interesting that the `correct' factor in Eq. \eqref{DecayRate} may be restored in the limit of the vanishingly small mass of the decaying particle: $m_0\ll m_1$. In this case, the equation \eqref{DecayRate2} becomes 
\begin{equation}
w\propto \exp\left(-\frac{2\pi m_1}{H}\right)= \exp\left(-\frac{m_1}{T_{\rm H}^{\rm dS}}\right)~.
\label{DecayRateHawking}
\end{equation}
The presence of the cosmological horizon does become important in this case, since in the limit  $m_0/m_1 \rightarrow 0$ the position $r_0$, to which the particle with mass $m_1$ is tunneling, approaches the horizon: $r_0 \rightarrow r_H$ when $m_0/m_1\rightarrow 0$.
One may argue that the limit, when the mass $m_0$ of the decaying particle approaches zero, formally corresponds to the creation of the pair of particles with mass $m_1$ from the vacuum; and this corresponds to the Hawking radiation from the de Sitter vacuum. 

However, this is not exactly true. The presence of the original particle is necessary for the radiation, otherwise the matrix element $\left<m_0|{\cal H}|m_1,m_1\right>$, which is  needed for the decay of the original particle \cite{Bros2009,Bros2008,Bros2006},  drops out. The presence of the original particle, even with zero mass, violates the symmetry of the de Sitter vacuum, and the radiation becomes possible (see discussion in Ref. \cite{Volovik2009}). Also one should not forget that the factor $2\pi$ in \eqref{DecayRateHawking} appears simply  because two particles  with masses $m_1$ tunnel simultaneously, that is why the effective activation  temperature, which appears in the decay of the composite particle in de Sitter background is $T=2T_{\rm H}^{\rm dS}$. It is still unclear whether there is a deep physics in this relation or just a coincidence.

Hawking radiation from the black hole can be also considered as the decay of the black hole with initial mass $M_{\rm i}$ concentrated in the singularity into a particle with the mass $m$  radiated  away and  the black hole with the smaller mass $M_{\rm f}<M_{\rm i}$. The corresponding trajectory of the radiated particle is 
\begin{equation}
E(p_r,r)   = \sqrt{m^2 + p_r^2}+ p_r v(r)= M_{\rm i}-M_{\rm f} ~.
\label{BH}
\end{equation}
If the black hole is immersed in the Minkowski spacetime, the energy conservation prescribes $M_{\rm i}=M_{\rm f}+m$ for particle radiated with zero momentum at infinity. Then the straightforward application of the semiclassical tunneling approach \cite{Volovik1999a} gives the radiation rate 
\begin{equation}
 w\propto \exp\left(-\frac{m}{T_{\rm H}^{\rm bh}}\right) ~,
\label{BH2}
\end{equation}
 with the correct Hawking temperature $T_{\rm H}^{\rm bh}$ for the black hole \cite{volovik2}.
This demonstrates that the black hole immersed in the Minkowski spacetime is decaying, while the Minkowski vacuum itself remains stable.  In the same manner the body (composite particle, atom, black hole or other object which serves as detector of radiation) immersed in the de Sitter vacuum is decaying, while the de Sitter vacuum may remain stable towards the Hawking radiation.

It is a pleasure to thank Vladimir Eltsov, Vincent Pasquier, Alexander Polyakov and Alexei Starobinsky for useful comments. This work is supported in part by the Russian Foundation
for Basic Research (grant 06--02--16002--a) and the
Khalatnikov--Starobinsky leading scientific school (grant
4899.2008.2).



\begin{thebibliography}{99}
\bibitem{volovik2}
If the black hole is immersed into the de Sitter vacuum, the radiation  with the  mass deficit, $M_{\rm i}<M_{\rm f} + m$, becomes possible. For this case the  ``fluid'' metric  in the form of Eq.\eqref{FluidMetric1} does not exist because of different sign of the frame dragging velocity $v(r)$ in \eqref{FRWfluid} and in \eqref{Schwarzschild}. Nevertheless, if the black hole and cosmological horizons are far apart, 
$T_{\rm H}^{\rm dS}\ll T_{\rm H}^{\rm bh}$, one may calculate the tunneling exponents separately. Then in the limit of small mass deficit,  $m+M_{\rm f}-M_{\rm i}\ll  m$, one obtains 
$ w(m,M_{\rm i}-M_{\rm f})\propto \exp\left(-\frac{m}{T_{\rm H}^{\rm bh}} 
-\frac{m+M_{\rm f}-M_{\rm i}}{2T_{\rm H}^{\rm dS}}\right)$.
\end{thebibliography}
\end{document}